# The preformation probability inside α-emitters having different ground state spin-parity than daughters

W. M. Seif, M. M. Botros, and A. I. Refaie
Cairo University, Faculty of Science,
Department of Physics, Giza 12613, Egypt

**ABSTRACT:**

The ground-state spin and parity of a formed daughter in the radioactive α-emitter is expected to influence the preformation probability of the α and daughter clusters inside it. We investigate the α and daughter preformation probability inside odd-A and doubly-odd radioactive nuclei when the daughter and parent are of different spin and/or parity. We consider only the ground-state to ground-state unfavored decays. This is to extract precise information about the effect of the difference in the ground states spin-parity of the involved nuclei far away any influences from the excitation energy if the decays are coming from isomeric states. The calculations are done for 161 α-emitters, with $65 \leq Z \leq 112$ and $84 \leq N \leq 173$, in the framework of the extended cluster model, with WKB penetrability and assault frequency. We used a Hamiltonian energy density scheme based on Skyrme-SLy4 interaction to compute the interaction potential. The α plus cluster preformation probability is extracted from the calculated decay width and the experimental half-life time. We discussed in detailed steps the effect of angular momentum of the emitted α-particle on the various physical quantities involved in the unfavored decay process and how it is finally increases the half-life time. We found that if the ground states spin and/or parity of parent and daughter nuclei are different, then the preformation probability of the α-cluster inside parent is less than it would be if they have similar spin-parity. We modified the formula that gives the α preformation probability in terms of the numbers of protons and neutrons outside the shell closures of parent, to account for this hindrance in the preformation probability for the unfavored decays between ground states.

## I. INTRODUCTION

In principle, the preformation probability of an emitted cluster inside the superheavy nucleus is one of the essential factors in predicting its dominated decay channel. Other factors are the released energy and the penetration probability. The uncertainties pertaining to the obtained values of the α-cluster preformation probability, or the α spectroscopic factor [1,2,3], in the different studies [4,5,6,7,8] make it one of the open problems. The accurate determination of the preformation probability would help in estimating the half-life times of the superheavy elements, suggested to synthesize in future, against their $\alpha$ decay modes. However, there are many confirmed factors to affect the α-cluster preformation probability. One of these factors is the deformation of the daughter that formed beside the α-cluster in the decaying nucleus, just before the decay process. The effect of the daughter deformation is found to decrease the preformation probability [7]. The second factor is the shell effect. As a function of the charge and neutron numbers, general oscillating patterns of the α preformation probability are observed with several local maxima and minima. While the local minima are observed at the proton and neutron shell and subshell closures, the coexisting maxima are predicted around mid-shell occupation numbers of the proton and neutron open shells. Third affecting factor is the isospin asymmetry of the parent nucleus [3]. It is clarified that the preformation probability increases with increasing the isospin asymmetry of the α emitters, if they have valence protons and neutrons but not holes. In case of coexisting neutron or proton holes with the valence ones, the preformation probability decreases upon increasing the isospin asymmetry. The preformation probability is shown to exhibit individual linear behaviors as a function of the isospin asymmetry parameter multiplied by the valence proton and neutron numbers. The obtained linear behaviors are correlated with the shell closures in the parent nuclei [3]. Also, existence of unpaired nucleons in the open shells of a parent nucleus influences the preformation process. This is the fourth affecting factor. Generally, the largest preformation probability of the α-cluster inside a parent is assigned for spherical even(Z)-even(N) α emitters, which have no unpaired nucleons. The α preformation probability in the decaying nuclei which have unpaired nucleons is shown to be less than it would be in their even-even neighboring isotopes, and isotones, of the same shell and subshell closures [9]. The effect of an unpaired neutron in hindering the preformation probability inside even-odd emitters is appeared to be slightly larger than that of the unpaired proton inside the odd-even nuclei. The smallest α preformation probability is predicted inside the deformed odd-odd decaying nuclei, which have an unpaired neutron and an unpaired proton at the same time.

Various approaches have been used to calculate the preformation probability. For instance, it can be calculated from the probability amplitude of the $\alpha$ and daughter clustering amount inside the parent nucleus [4], or from the formation energy of the α-cluster extracted from the binding energy differences of the involved nuclei [10]. To calculate the formation amplitude and the α decay probability by applying the *R*-matrix approach, the Skyrme-Hartree-Fock-Bogoliubov wave functions have been used in several studies [11]. Also, the $\alpha$ particle formation probabilities are deduced using the universal decay law method based on the *R*-matrix [6]. Furthermore, it can be extracted from its explicit relation to the experimental half-life time [2,3,12] and the calculated decay width.

Concerning the unfavored decay process, one of the major consequences of the difference between the ground state spin-parity of parent and daughter is the angular momentum carried away by the emitted $\alpha$-particle. Indeed, the general trend of increasing the half-life time for such unfavored α decays due to the angular momentum transferred by the emitted α-particle is considered in several studies [13,14]. Now, the pivotal question arises whether the preformation of the α plus daughter clusters have an independent probability regardless of the ground state spin-parity ($J^{\pi}$) of the parent and daughter. That is to say, if the parent and the formed daughter are of different spin-parity ground states, will the preformation probability influenced? We want to concern this question in the present work.

We organize the paper as follows. In the next section, the theoretical approach of extracting and investigating the α preformation probability is outlined. The results for the studied nuclei are discussed in Sec. III. Finally, Sec. IV presents a brief summary and conclusion.

## II. THEORETICAL FORMALISM

In the preformed cluster approaches [15,16], the partial half-life time ($T_{1/2}$) for a specific decay mode of an $\alpha$ emitter in a given spin(J)-Parity($\pi$) ground state ($J_P^{\pi}$) leaving a daughter in a state ($J_D^{\pi}$) is given as,

$$T_{1/2} = \frac{ln\, 2}{S_\alpha \Gamma}. \tag{1}$$

Here, $S_\alpha$ denotes the preformation probability of the α and daughter as two individual clusters inside the decaying parent nucleus. $\Gamma$ is the decay width. For the decays involving deformed daughter nuclei, the decay width at a given orientation angle ($\theta$) reads,

$$\Gamma(\theta) = \nu(\theta) P(\theta). \tag{2}$$

$\theta$ is usually defined as the relative angle between the separation vector $\vec{r}$ joining the centers of mass of α and daughter and the symmetry-axis of the deformed daughter. In the framework of the well-known Wentzel-Kramers-Brillouin (WKB) approach, the tunneling assault frequency ($\nu(\theta)$) and penetration probability ($P(\theta)$) of the emitted $\alpha$-particle at a certain orientation $\theta$ are given, respectively, as

$$\nu(\theta) = T^{-1}(\theta) = \left[ \int_{R_1(\theta)}^{R_2(\theta)} \frac{2\mu}{\hbar k(r,\theta)} dr \right]^{-1}, \tag{3}$$

and

$$P(\theta) = e^{-2 \int_{R_2(\theta)}^{R_3(\theta)} k(r,\theta) dr}. \tag{4}$$

In terms of the reduced mass of the α ($m_\alpha$)-daughter ($m_D$) system ($\mu = \frac{m_\alpha m_D}{m_\alpha + m_D}$) and the experimental Q-value ($Q_\alpha$(MeV)) of the decay [17], the wave number $k(r,\theta)$ reads,

$$k(r,\theta) = \sqrt{\frac{2\mu}{\hbar^2}|V_T(r,\theta) - Q_\alpha|}\,. \tag{5}$$

Along the oscillating and tunneling path of the α particle, the three classical turning points $R_{i=1,2,3}$(fm) satisfy the condition, $V_T(r,\theta)|_{r=R_i(\theta)} = Q_\alpha$. While the first two turning points ($R_{1,2}(\theta)$) depend on the emitting orientation, $R_3$ is independent of it [9]. Three contributions are forming the total real interaction potential ($V_T(r,\theta)$) between the two interaction clusters inside and outside the parent nucleus. Namely, they are the nuclear ($V_N(r,\theta)$), Coulomb ($V_C(r,\theta)$) and the centrifugal ($V_\ell(r)$) parts,

$$V_T(r,\theta) = V_N(r,\theta) + V_C(r,\theta) + V_\ell(r). \tag{6}$$

We shall use the energy density formalism [18], with the frozen density approximation, to calculate nuclear interaction part in terms of an appropriate Skyrme interaction [19,20,21],

$$\begin{aligned} V_N(r,\theta) = \int \{ & H[\rho_{p\alpha}(\vec{x}) + \rho_{pD}(r,\vec{x},\theta), \rho_{n\alpha}(\vec{x}) + \rho_{nD}(r,\vec{x},\theta)] \\ & - H_\alpha[\rho_{p\alpha}(\vec{x}), \rho_{n\alpha}(\vec{x})] - H_D[\rho_{pD}(\vec{x}), \rho_{nD}(\vec{x})]\}\, d\vec{x}\,. \end{aligned} \tag{7}$$

The Skyrme energy-density functionals of the whole system (*H*), $\alpha$ ($H_\alpha$) and daughter ($H_D$) are given as functions of the proton (*p*) and neutron (*n*) frozen density distributions in the α and daughter clusters, $\rho_{ij}(i = p,n\ \&\ j = \alpha, D)$. The proton and neutron density distributions of the deformed nuclei are used in their two-parameter Fermi shape with radii and diffuseness parameters are adjusted to reproduce the density distributions obtained from the self-consistent Hartree-Fock calculations [9]. The quadrupole ($\beta_2$), octupole ($\beta_3$), hexadecapole ($\beta_4$) and hexacontatetrapole ($\beta_6$) deformations [22] are expressed in the half-density radii of the Fermi density function. The SLy4 parameterization of the Skyrme-like nucleon-nucleon (NN) force [23] is used in the present calculations. The Skyrme-like force has the advantage to include explicitly the pairing and shell effect influences in the calculations [9]. The nuclear potential is usually normalized with a factor λ by applying the Bohr-Sommerfeld quantization condition [24] to ensure a quasi-stationary state [25]. The Coulomb interaction ($V_C(r,\theta)$) is also calculated based on the direct and exchange Coulomb functionals [26],

$$\begin{aligned} H_{Coul}(\rho_p) &= H_C^{dir}(\rho_p) + H_C^{exch}(\rho_p) \\ &= \frac{e^2}{2}\rho_p(\vec{r})\int \frac{\rho_p(\vec{r}')}{|\vec{r}-\vec{r}'|}d\vec{r}' - \frac{3e^2}{4}\left(\frac{3}{\pi}\right)^{1/3}\left(\rho_p(\vec{r})\right)^{4/3}. \end{aligned} \tag{8}$$

To simplify the complications coming from the finite range of the proton-proton Coulomb force, the density multipole expansion of the deformed daughter has been used to calculate the Coulomb direct part [27,28]. If the ground states spin-parity of parent ($J_P^\pi$) and daughter ($J_D^\pi$) are not identical, we have then unfavored decay mode. In such cases, the centrifugal potential part reads,

$$V_\ell(r) = \frac{\ell(\ell+1)\hbar^2}{2\mu r^2}. \tag{9}$$

$\ell$ here stands for the angular momentum transferred by the emitted $\alpha$-particle to conserve the spin and parity in the decay process. As the ground state spin-parity of the α-particle is $0^+$, the spin and parity conservation laws for the decay yields,

$$|J_P - J_D| \le \ell \le |J_P + J_D|, \tag{10}$$

and

$$\pi_P = \pi_D(-1)^\ell. \tag{11}$$

According to the principle of least action, we shall consider the minimum value of $\ell$ verifying Eqs. (10) and (11) as the preferred angular momentum carried out by the $\alpha$ particle. In this sense, $\ell_{min}$ will be fixed as the minimum even (odd) value of $\ell$ from Eq. (10) if $\pi_P$ and $\pi_D$ are similar (different).

However, by the orientation averaging we obtain the average decay width as,

$$\Gamma = \frac{1}{2}\int_0^\pi \Gamma(\theta)\, sin\,\theta\ d\theta. \tag{12}$$

We can now debrief the preformation probability of the $\alpha$-cluster in the $\alpha$ emitter from the experimental half-life time and the calculated decay width as,

$$S_\alpha = \frac{ln\,2}{\Gamma\, T_{1/2}^{exp}}. \tag{13}$$

Based on its observed behaviour with $Z$ and $N$, a semi-empirical formula was proposed [7,9] to give the $\alpha$ preformation probability in terms of the numbers of protons $(Z - Z_0)$ and neutrons $(N - N_0)$ outside the shell and subshell closures $(Z_0, N_0)$ in the $\alpha$ emitter,

$$S_\alpha = Ae^{-\alpha(Z-Z_0-Z_c)^2} e^{-\beta(N-N_0-N_c)^2} - a_p$$

$$a_p = \begin{cases} 0.0040\,(Z-Z_0)^{\frac{1}{3}} & \text{for odd}(Z)-\text{even}(N)\text{ nuclei} \\ 0.0056\,(N-N_0)^{\frac{1}{3}} & \text{for even}(Z)-\text{odd}(N)\text{ nuclei} \\ 0.0088\,(Z-Z_0+N-N_0)^{\frac{1}{3}} & \text{for odd}(Z)-\text{odd}(N)\text{ nuclei.} \end{cases} \tag{14}$$

Here $Z_c(N_c)$ represents the number of protons (neutrons), outside the $Z_0(N_0)$ shell closures, which yields a local maximum value for the preformation probability. $A$, $\alpha$ and $\beta$ are dimensionless parameters. The pairing term $(a_p)$ takes account for the influences of the unpaired nucleon(s), inside the odd-A and odd(Z)-odd(N) α emitters, on the $\alpha$-preformation probability [9].

## III. RESULTS AND DISCUSSION

As mentioned above, our study is confined to the ground state to ground state decays to scrutinize the mere effect of the difference in the ground states spin-party, excluding any other effects due the excitation energy for the decays from isomeric states. The decays from 161 open-shell odd-A and odd (Z)-odd(N) radioactive α emitters in the mass region of $A = 149 - 285$ are mentioned. In Table I we list related structure information on the investigated $\alpha$ decays and the deduced preformation probabilities, $S_{\alpha}^{exp}$ (Eq. (13) ). The experimental errors in both the Q-value [17], column 6, and the experimental half-live [29-36], column 7, are taken into account in deducing the preformation probability. The first five columns of Table I identify, respectively, the parent and daughter nuclei, and the ground-state spin and parity of them ($J_{P(D)}^{\pi}$) in addition to the considered value of minimum angular momentum carried out by the emitted α particle ($\ell_{min}$). Presented in the ninth and tenth columns, respectively, are the estimated preformation probabilities using the modified semi-empirical formula given by Eq.(15), mentioned below, and the half-life times as obtained based on these values. Table I shows that only eight nuclei (5%) out of the 161 investigated α emitters yield a preformation probability greater than 0.1. The predicted value of $S_{\alpha}^{exp}$ is in the order of $10^{-2}$ and $10^{-3}$ for seventy (≈ 44%) and sixty (≈ 37%) nuclei, respectively. Twenty two studied cases (≈ 14%) yield a smaller $S_{\alpha}^{exp}$ in the order of $10^{-4}$. The α preformation probability in one of the investigated emitters ($^{244}$Bk) is obtained with a value less than 0.0001. In the similar study has been performed on the favored decays of 105 odd-A and odd(Z)-odd(N) nuclei [9], 6% , 80%, 13% and only one nucleus among the investigated nuclei yielded α preformation probability in the order of $10^{-1}$ , $10^{-2}$ , $10^{-3}$ and $10^{-4}$, respectively. Larger relative number of α emitter (25%) yielding preformation probability greater than 0.1 is obtained for the studied even-even nuclei in Ref. [7]. The average α preformation probability inside all odd-even, even-odd and odd-odd parent nuclei participating in the unfavored decays mentioned in the present work are 0.0256, 0.0303 and 0.0161, respectively. The corresponding average values inside the parent nuclei involved in the favored decays mentioned in in Ref. [9] are 0.0446, 0.0361 and 0.0322, respectively. However, the preformation probability inside odd-A and odd-odd nuclei exhibits relative smaller values for unfavored decays with respect to the favored ones.

Beside an expected change in the α-preformation probability involved in the unfavored decays, the non-zero angular momentum carried by the emitted α particle impacts the different parameters and quantities participating in the decay process. The results show that, as $\ell$ increases, the width of the internal pocket of the total interaction potential ($R_2 - R_1$) decreases and its lowest point shifts up. These two factors increase the assault frequency. On the other hand, the increasing of $\ell$ produces higher Coulomb barrier characterized with a wider barrier width, $R_3 - R_2$. Also, the radius of the Coulomb barrier slightly decreases. Subsequently, the penetration probability deceases. The relative reduction in the penetration probability due to the non-zero $\ell$ is more prominent, in the decay process, than the increase in the assault frequency. Consequently, the net effect of these two opposite factors is to hinder the decay width. However, the overall impact of the non-zero transferred angular momentum on the unfavored decay mode appears as a substantial increase in half-life time. Of course, the presence of barrier distribution due to any deformation in the involved nuclei increases the range of variation of all mentioned quantities and consequently the effect of $\ell$ appears more clearly.

Displayed in Fig. 1 are the estimated α preformation probability in the even-odd isotopes of Rn (Z=86) and Cm (Z=96), the odd-even isotopes of Ac (Z=89) and Pa (Z=91), and the odd-odd ones of At(Z=85) and Pa (Z=91), as a function of the neutron number of the α emitters. In this figure, the present results for the unfavored decays with $\ell_{min} \neq 0$ (open symbols) are compared with those of the favored decays with $\ell_{min} = 0$ (solid symbols) for the same mentioned isotopic chains, taken from Ref. [9]. Generally, the presented results follow the oscillatory behavior detected for $S_\alpha$ as a function of N [7]. The average $S_\alpha^{exp}$ for the shown unfavored decays involving odd-even, even-odd and odd-odd are 0.0151, 0.0176 and 0.0161, respectively. The corresponding values for the displayed favored decays are, respectively, 0.0486, 0.0559 and 0.0270. Moreover, there is large uncertainty in the high appearing values of $S_\alpha^{exp}$ for the unfavored decays of $^{205}Rn(0.0960 \pm 0.0488)$ and $^{220}At(0.0935 \pm 0.0421)$. These large uncertainties are due to large ones in the released energy value for the decay of $^{205}Rn(Q_\alpha = 6.390 \pm 0.050)$ and in the experimental half-life time and the ground-state spin and parity of the involved nuclei for the decay of $^{220}At$, as shown in Table I. In Fig. 2, the extracted preformation probability inside the even-odd isotones of N=121,133 and N=157, and the odd-even isotones of N=126, 130 and154, in addition to the odd-odd ones of N=129,131 and 133 are presented versus the charge number of the α emitters. The results obtained in the present work for the unfavored decays are compared in the same figure with the corresponding values for the favored decays which are obtained in Ref. [9]. For the unfavored decays, the deduced preformation probability inside the odd-even isotones presented in Fig. 2 ranges between $0.0002 \pm 0.0001$ ($^{255}$Md) and $0.0108 \pm 0.0062$ ($^{259}$Db). Inside the presented even-odd (odd-odd) isotones, the extracted $S_\alpha^{exp}$ has values lying between $0.0040 \pm 0.0001$($^{257}$Fm) and $0.0761 \pm 0.0480$($^{211}$Th) (0.0014($^{220}$Ac) and $0.0296 \pm 0.0008$ ($^{220}$Fr)). The corresponding values for the presented favored decays of odd-even, even-odd and odd-odd nuclei lie between $0.0069 \pm 0.0001$($^{211}$At) and $0.1180 \pm 0.0485$($^{219}$Ac), $0.0044 \pm 0.0022$($^{263}$Sg) and $0.0527 \pm 0.0018$ ($^{217}$Po), and $0.0299 \pm 0.0069$($^{218}$At) and $0.0692 \pm 0.0421$($^{218}$Fr), respectively. In general, the average preformation probability for the displayed unfavored (favored) decays of odd-even, even-odd and odd-odd nuclei are 0.0045(0.0481), 0.0242(0.0280) and 0.0119(0.0453), respectively. Again, the results presented in Figs. 1 and 2 indicate that the preformation probability tends to decrease if the parent and daughter nuclei have different ground-state spins and/or parities.

Now, in context of our results, we shall modify the semi-empirical formula given by Eq. (14) to account for the hindrance in the preformation probability in the unfavored decays, with respect to the favoured ones. We suggest the modified expression to read,

$$S_\alpha = \frac{Ae^{-\alpha(Z-Z_0-Z_c)^2} e^{-\beta(N-N_0-N_c)^2} - a_p}{a_\ell}. \tag{15}$$

Here, the coefficient $a_\ell$ is introduced to account for the difference between the ground states spin and/or parity of the participating nuclei. However, for favoured decays, $a_\ell = 1$. Our next step is to do two things. First, we try to reduce the number of fit parameters of Eq. (15) by fixing some of its variable parameters given in Refs. [7,9]. Second, we look into the form of $a_\ell$. By optimizing the fit to the results of 284 even-even, odd-A and odd-odd favored decays given in Refs.[7,9] and keeping a fixed values of $\alpha = 0.003$ and $\beta = 0.006$, the parameterization of Eq. (15) is listed in Table II. The paring term ($a_p$) is still given by Eq. (14). With respect to the experimental half-life times of the mentioned 284 favoured decays, taking account of the experimental errors, the standard deviation of their calculated half-lives based the parameterization of $S_\alpha$(Eq. (14)) given in Table II is $\sigma = 0.375$. The standard deviation is given as, $\sigma = \sqrt{\sum_{i=1}^{n}\left[log_{10}\left(T_{1/2}^{cal}/T_{1/2}^{exp}\right)\right]^2/(n-1)}$. Based on the present results of $S_\alpha^{exp}$ for the studied 161 unfavored decays, the $a_\ell$ coefficient is adopted to be given in terms of the mass number of the parent nucleus (A) and the minimum allowed value of the transferred angular momentum ($\ell_{min}$) as,

$$a_\ell = \frac{A^2}{3190\sqrt{\ell_{min}}} - 3.9\ . \tag{16}$$

Considering the hindrance in preformation probability due to the difference in the ground states spin-parity of the involved nuclei improves substantially the standard deviation of the half-life times calculated using $S_\alpha$(Eq. (15)) from $\sigma = 1.139$ (without $a_\ell$) to $\sigma = 0.725$. The calculated half-lives of $^{236}Np(S_\alpha^{exp} = 0.0001 \pm 0.0001)$, $^{244}Bk$ $(S_\alpha^{exp} = 0.00001 \pm 0.00001)$ and $^{228}Pa(S_\alpha^{exp} = 0.0003)$ yield the largest deviations from the experimental ones. Actually, in addition to the low extracted preformation probability inside these three decays, they are characterised with α-decay mode of low intensity [29]. Moreover, the available experimental half-lives and intensities on these decays are little old [29,37,38]. If we neglect these three decays, the obtained standard deviation becomes $\sigma$ =0.667. For all mentioned 445 decays, the total standard deviation of the calculated half-life times, based on the preformation probability given by Eq. (15), from the experimental values is $\sigma$ =0.530.

## IV. SUMMARY AND CONCLUSION

We studied the ground-state to ground-state unfavored decays of 161 odd(Z)-even(N), even-odd and odd-odd open-shell nuclei. We investigated the preformation probability inside these nuclei. We also explored the role of the transferred angular momentum by the emitted α particle in the different physical quantities involved in the unfavored decay process. The detailed calculations showed that the angular momentum of the emitted α particle increases the assault frequency where it decreases the width of the internal pocket of the interaction potential shifting its bottom up. Conversely but more effectively, it reduces the penetration probability because it produces higher Coulomb barrier with wider barrier width. The net effect appears as hindrance in the decay width. Consequently, the half-life increases substantially for the unfavored decays. Most important, we found that for a parent with a given ground-state spin-parity, the probability of forming an α-cluster and a daughter nucleus with different ground-state spin and/or parity is less than it if the daughter nucleus has the same spin and parity of parent. The formula relating the α preformation probability to the numbers of protons and neutrons outside the closed shells in the parent nuclei is modified to take account of the hindrance in the preformation probability associated with the unfavored decays between ground states.

## Figures and Tables captions:

**Fig. 1**: The estimated α preformation probability (Eq.(13)) inside even(Z)-odd(N) Rn and Cm isotopes, odd-even Ac and Pa isotopes, and odd-odd At and Pa isotopes versus the neutron number of parent α emitter. While the open symbols represent unfavored decays with involved transferred angular momentum ($\ell_{\min}$), the solid ones represent favored decays ($\ell_{\min}$=0) taken from Ref. [9].

**Fig. 2**: Same as Fig. 1 but for the even-odd isotonic chains of N=121,133,157, the odd-even isotonic chains of N=126,130,154 and the odd-odd isotonic chains of N=129,131,133 versus the charge number of parent α emitters.

**Table I**. The estimated α preformation probability, $S_{\alpha}^{exp}$ (Eq.(13)), in the mentioned parent nuclei based on the experimental partial half-lives ($T_{1/2}^{exp}$ (s)) [29-36]. The uncertainties in the experimental half-life time and the intensity are both considered in the listed partial half-lives. The decay width is calculated using WKB penetration probability and assault frequency, based on Skyrme-SLy4 NN interaction. The first six columns identify, respectively, the parent and daughter nuclei, their ground-state spin and parity ($J_{P(D)}^{\pi}$), the minimum expected value of the angular momentum carried out by the emitted α particle ($\ell_{min}$), and the energy released through the decay process ($Q_{\alpha}$-values) [17]. The non-experimental estimated spin and/or parity and their uncertain values are indicated in square and curve brackets [29], respectively. The last two columns exhibit the estimated preformation probability from the formula given by Eq. (15) and the calculated partial half-lives based on these values, respectively.

**Table II**: The fit parameters of the semi-empirical expression given by Eq. (15) for the α preformation probability inside 284 favored decays [7,9]. The values of the dimensionless parameters $\alpha$ and $\beta$ are kept fixed at $\alpha = 0.003$ and $\beta = 0.006$.

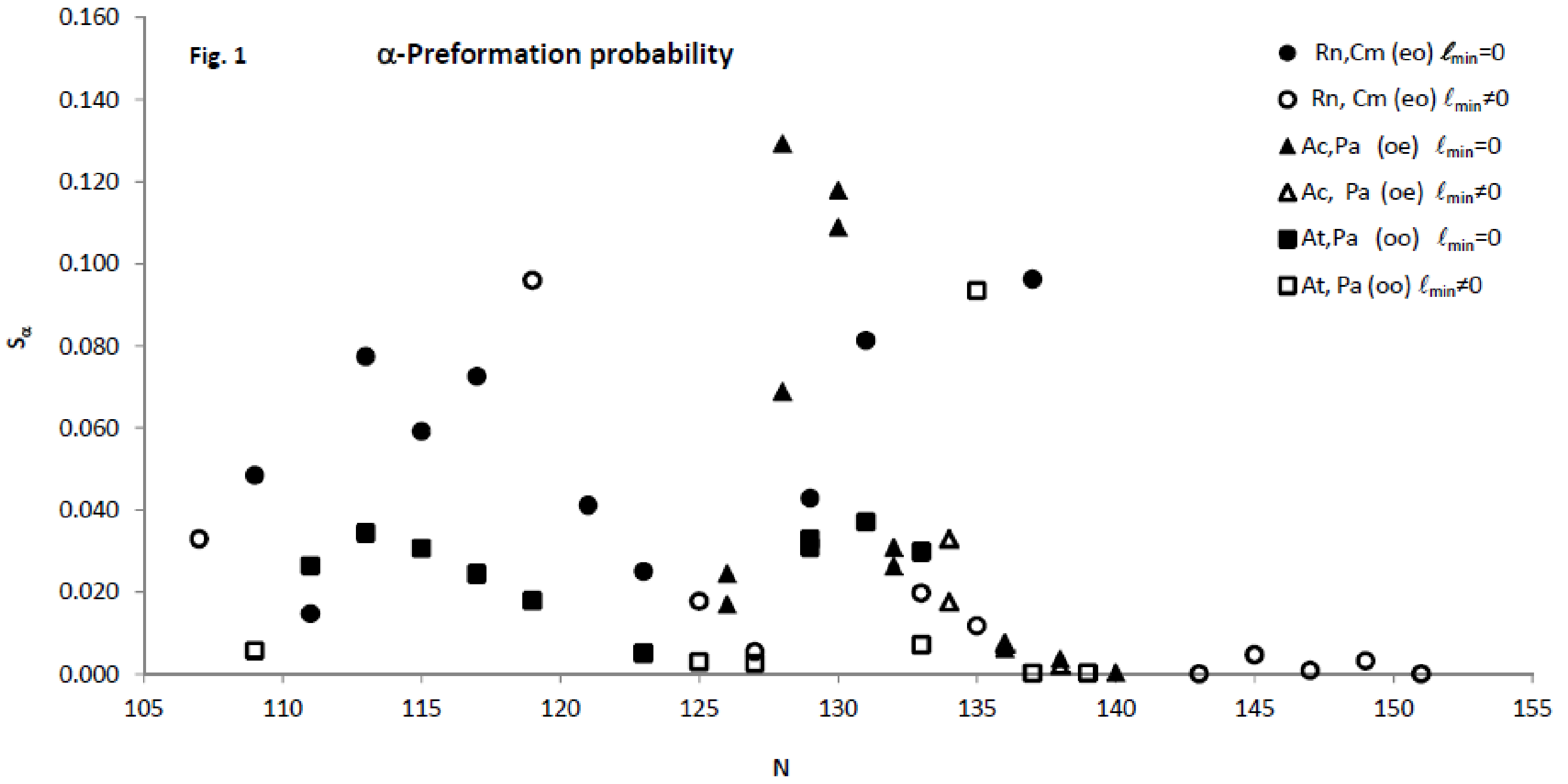

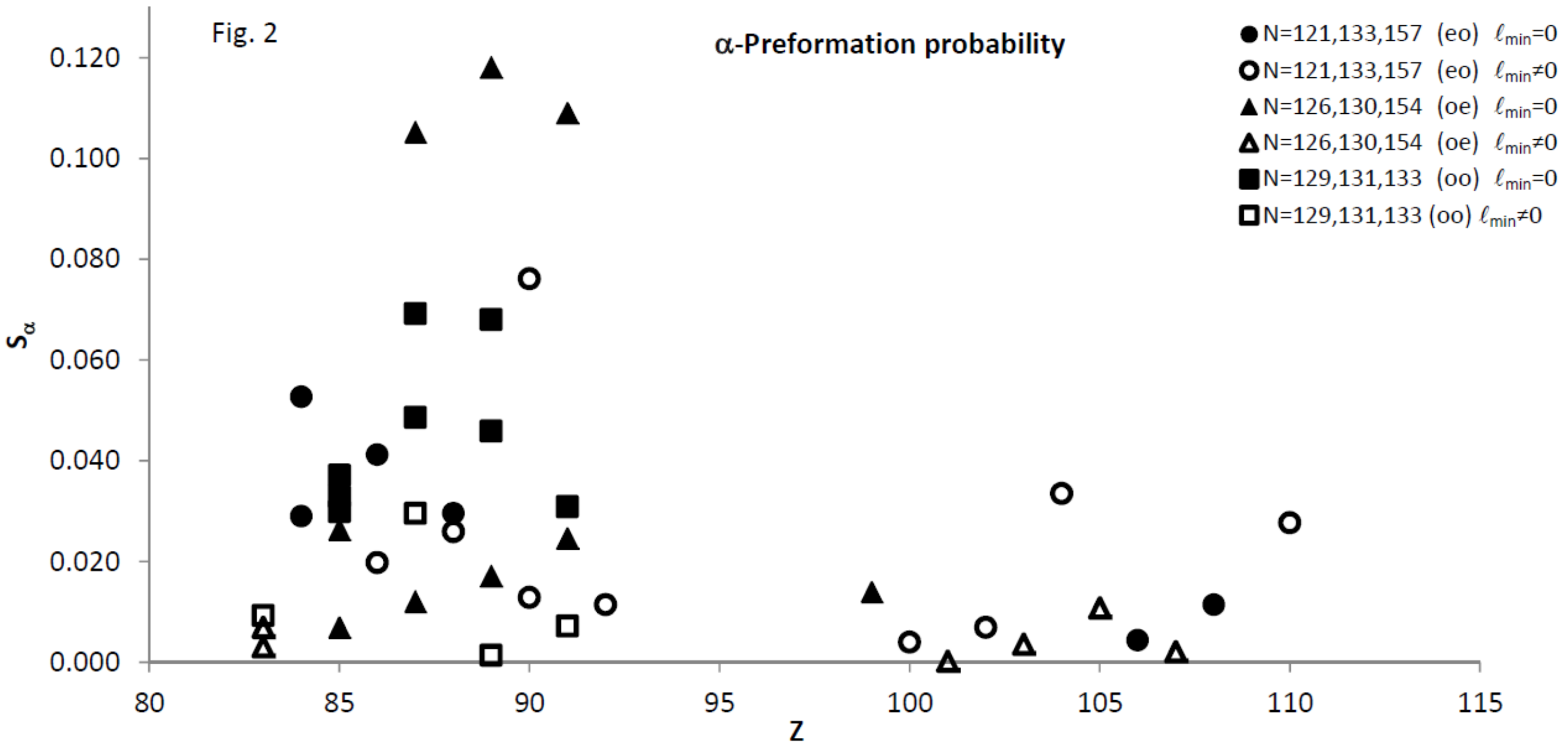

**Table I**.

| Pare. | Dau. | $J_P^\pi$ | $J_D^\pi$ | $\ell_m$ | $Q_\alpha$ (MeV) | $T_{1/2}^{exp}$ (s) | $S_\alpha^{exp}$ (Eq. (13)) | $S_\alpha$ ($Eq.$ (15)) | $T_{1/2}^{cal}$ (s) (using $S_\alpha$ (Eq. (15))) |
|---|---|---|---|---|---|---|---|---|---|
| $^{149}Tb$ | $^{145}Eu$ | $1/2^+$ | $5/2^+$ | 2 | $4.078 \pm 0.002$ | $(8.976 \pm 0.968) \times 10^4$ | $0.0311 \pm 0.0044$ | 0.0650 | $4.315^{+0.055}_{-0.232} \times 10^4$ |
| $^{151}Tb$ | $^{147}Eu$ | $1/2^{(+)}$ | $5/2^+$ | 2 | $3.497 \pm 0.004$ | $(6.844 \pm 1.081) \times 10^8$ | $0.0526 \pm 0.0125$ | 0.0660 | $5.176^{+0.514}_{-0.358} \times 10^8$ |
| $^{173}Hg$ | $^{169}Pt$ | $[3/2^-]$ | $(7/2^-)$ | 2 | $7.378 \pm 0.004$ | $(9.100 \pm 2.600) \times 10^{-4}$ | $0.1164 \pm 0.0365$ | 0.0181 | $54.330^{+0.452}_{-2.767} \times 10^{-4}$ |
| $^{177}Hg$ | $^{173}Pt$ | $(7/2^-)$ | $(5/2^-)$ | 2 | $6.740 \pm 0.050$ | $(1.498 \pm 0.021) \times 10^{-1}$ | $0.0709 \pm 0.0294$ | 0.0117 | $8.241^{+4.463}_{-2.833} \times 10^{-1}$ |
| $^{181}Hg$ | $^{177}Pt$ | $1/2^{[-]}$ | $5/2^-$ | 2 | $6.284 \pm 0.004$ | $13.434 \pm 1.366$ | $0.0143 \pm 0.0022$ | 0.0053 | $35.161^{+2.320}_{-1.339}$ |
| $^{180}Tl$ | $^{176}Au$ | $4^{(-)}$ | $(5^-)$ | 2 | $6.710 \pm 0.050$ | $32.900 \pm 22.100$ | $0.0024 \pm 0.0021$ | 0.0041 | $6.117^{+5.777}_{-1.090}$ |
| $^{179}Pb$ | $^{175}Hg$ | $(9/2^-)$ | $(7/2^-)$ | 2 | $7.598 \pm 0.020$ | $3.500^{+1.400}_{-0.800} \times 10^{-3}$ [31] | $0.0230 \pm 0.0096$ | 0.0073 | $10.503^{+1.630}_{-1.403} \times 10^{-3}$ |
| $^{181}Pb$ | $^{177}Hg$ | $(9/2^-)$ | $(7/2^-)$ | 2 | $7.240 \pm 0.007$ | $(3.900 \pm 0.080) \times 10^{-2}$ | $0.0259 \pm 0.0019$ | 0.0049 | $20.414^{+1.158}_{-1.056} \times 10^{-2}$ |
| $^{183}Pb$ | $^{179}Hg$ | $3/2^-$ | $7/2^-$ | 2 | $6.928 \pm 0.007$ | $0.535 \pm 0.030$ | $0.0237 \pm 0.0028$ | 0.0028 | $4.476^{+0.311}_{-0.272}$ |
| $^{185}Pb$ | $^{181}Hg$ | $3/2^-$ | $1/2^{[-]}$ | 2 | $6.695 \pm 0.005$ | $6.300 \pm 0.400$ | $0.0107 \pm 0.0011$ | 0.0053 | $12.582^{+0.480}_{-0.574}$ |
| $^{187}Pb$ | $^{183}Hg$ | $3/2^-$ | $1/2^-$ | 2 | $6.393 \pm 0.006$ | $(1.681 \pm 0.386) \times 10^2$ | $0.0057 \pm 0.0016$ | 0.0063 | $1.428^{+0.094}_{-0.077} \times 10^2$ |
| $^{189}Pb$ | $^{185}Hg$ | $3/2^-$ | $1/2^-$ | 2 | $5.870 \pm 0.040$ | $(1.263 \pm 0.053) \times 10^4$ | $0.0136 \pm 0.0062$ | 0.0073 | $2.061^{+1.221}_{-0.713} \times 10^4$ |
| $^{184}Bi$ | $^{180}Tl$ | $[3^+]$ | $4^{(-)}$ | 1 | $8.020 \pm 0.050$ | $(6.600 \pm 1.500) \times 10^{-3}$ | $0.0011 \pm 0.0006$ | 0.0003 | $17.985^{+7.247}_{-4.998} \times 10^{-3}$ |
| $^{186}Bi$ | $^{182}Tl$ | $(3^+)$ | $(7^+)$ | 4 | $7.757 \pm 0.012$ | $(1.480 \pm 0.070) \times 10^{-2}$ | $0.0167 \pm 0.0022$ | 0.0105 | $2.344^{+0.213}_{-0.193} \times 10^{-2}$ |
| $^{187}Bi$ | $^{183}Tl$ | $[9/2^-]$ | $1/2^{(+)}$ | 5 | $7.779 \pm 0.004$ | $0.037 \pm 0.002$ | $0.0168 \pm 0.0014$ | 0.0256 | $0.024^{+0.001}_{-0.001}$ |
| $^{188}Bi$ | $^{184}Tl$ | $[3^+]$ | $[2^-]$ | 1 | $7.264 \pm 0.005$ | $(6.120 \pm 0.270) \times 10^{-2}$ | $0.0202 \pm 0.0013$ | 0.0028 | $43.420^{+0.917}_{-0.705} \times 10^{-2}$ |
| $^{189}Bi$ | $^{185}Tl$ | $(9/2^-)$ | $[1/2^+]$ | 5 | $7.268 \pm 0.003$ | $0.658 \pm 0.047$ | $0.0415 \pm 0.0039$ | 0.0297 | $0.918^{+0.019}_{-0.024}$ |
| $^{190}Bi$ | $^{186}Tl$ | $(3^+)$ | $(2^-)$ | 1 | $6.863 \pm 0.004$ | $6.300 \pm 0.100$ | $0.0046 \pm 0.0002$ | 0.0034 | $8.357^{+0.285}_{-0.044}$ |
| $^{191}Bi$ | $^{187}Tl$ | $(9/2^-)$ | $(1/2^+)$ | 5 | $6.778 \pm 0.003$ | $12.400 \pm 0.300$ | $0.1239 \pm 0.0079$ | 0.0326 | $46.587^{+1.962}_{-1.042}$ |
| $^{192}Bi$ | $^{188}Tl$ | $(3^+)$ | $(2^-)$ | 1 | $6.376 \pm 0.005$ | $(3.527 \pm 1.545) \times 10^2$ | $0.0077 \pm 0.0036$ | 0.0040 | $5.425^{+0.238}_{-0.248} \times 10^2$ |
| $^{193}Bi$ | $^{189}Tl$ | $(9/2^-)$ | $(1/2^+)$ | 5 | $6.304 \pm 0.005$ | $(2.271 \pm 1.059) \times 10^3$ | $0.0705 \pm 0.0356$ | 0.0342 | $3.575^{+0.187}_{-0.173} \times 10^3$ |
| $^{194}Bi$ | $^{190}Tl$ | $(3^+)$ | $2^{(-)}$ | 1 | $5.918 \pm 0.005$ | $(2.981 \pm 1.685) \times 10^4$ | $0.0104 \pm 0.0062$ | 0.0043 | $4.778^{+0.267}_{-0.190} \times 10^4$ |
| $^{195}Bi$ | $^{191}Tl$ | $(9/2^-)$ | $(1/2^+)$ | 5 | $5.832 \pm 0.005$ | $(1.114 \pm 0.756) \times 10^6$ | $0.0249 \pm 0.0178$ | 0.0342 | $0.414^{+0.033}_{-0.025} \times 10^6$ |
| $^{196}Bi$ | $^{192}Tl$ | $(3^+)$ | $2^{(-)}$ | 1 | $5.440 \pm 0.040$ | $(2.950 \pm 0.976) \times 10^7$ | $0.0023 \pm 0.0015$ | 0.0043 | $1.049^{+0.683}_{-0.390} \times 10^7$ |
| $^{209}Bi$ | $^{205}Tl$ | $9/2^-$ | $1/2^+$ | 5 | $3.137 \pm 0.001$ | $(6.280 \pm 0.221) \times 10^{26}$ | $0.0031 \pm 0.0002$ | 0.0120 | $1.640^{+0.023}_{-0.053} \times 10^{26}$ |
| $^{211}Bi$ | $^{207}Tl$ | $9/2^-$ | $1/2^+$ | 5 | $6.750 \pm 0.001$ | $128.400 \pm 1.200$ | $0.0045 \pm 0.0001$ | 0.0154 | $37.833^{+0.189}_{-0.221}$ |
| $^{212}Bi$ | $^{208}Tl$ | $1^{(-)}$ | $5^+$ | 5 | 6.207 | $(1.011 \pm 0.003) \times 10^4$ | 0.0093 | 0.0133 | $0.708 \times 10^4$ |
| $^{213}Bi$ | $^{209}Tl$ | $9/2^-$ | $(1/2^+)$ | 5 | $5.982 \pm 0.006$ | $(1.309 \pm 0.021) \times 10^5$ | $0.0069 \pm 0.0005$ | 0.0186 | $0.485^{+0.030}_{-0.029} \times 10^5$ |
| $^{189}Po$ | $^{185}Pb$ | $(5/2^-)$ | $3/2^-$ | 2 | $7.694 \pm 0.015$ | $(3.800 \pm 0.400) \times 10^{-3}$ | $0.0691 \pm 0.0146$ | 0.0071 | $36.011^{+4.163}_{-3.697} \times 10^{-3}$ |
| $^{203}Po$ | $^{199}Pb$ | $5/2^-$ | $3/2^-$ | 2 | $5.496 \pm 0.005$ | $(2.075 \pm 0.405) \times 10^6$ | $0.0588 \pm 0.0160$ | 0.0071 | $15.914^{+1.677}_{-0.959} \times 10^6$ |
| $^{209}Po$ | $^{205}Pb$ | $1/2^-$ | $5/2^-$ | 2 | $4.979 \pm 0.001$ | $(3.219 \pm 0.158) \times 10^9$ | $0.0222 \pm 0.0013$ | 0.0024 | $29.434^{+0.326}_{-0.318} \times 10^9$ |
| $^{211}Po$ | $^{207}Pb$ | $(9/2^+)$ | $1/2^-$ | 5 | $7.595 \pm 0.001$ | $0.516 \pm 0.003$ | 0.0033 | 0.0140 | $0.121^{+0.000}_{-0.001}$ |
| $^{194}At$ | $^{190}Bi$ | $(4^-, 5^-)$ | $(3^+)$ | 1 | $7.462 \pm 0.015$ | $0.253 \pm 0.010$ | $0.0058 \pm 0.0009$ | 0.0044 | $0.325^{+0.040}_{-0.034}$ |
| $^{195}At$ | $^{191}Bi$ | $1/2^+$ | $(9/2^-)$ | 5 | $7.339 \pm 0.005$ | $0.328 \pm 0.020$ | $0.2725 \pm 0.0296$ | 0.0347 | $2.538^{+0.142}_{-0.103}$ |
| $^{210}At$ | $^{206}Bi$ | $(5)^+$ | $6^{(+)}$ | 2 | $5.631 \pm 0.001$ | $(1.698 \pm 0.276) \times 10^7$ | $0.0030 \pm 0.0006$ | 0.0009 | $5.548^{+0.064}_{-0.163} \times 10^7$ |
| $^{212}At$ | $^{208}Bi$ | $(1^-)$ | $5^+$ | 5 | $7.817 \pm 0.001$ | $0.314 \pm 0.002$ | $0.0027 \pm 0.0001$ | 0.0112 | $0.075^{+0.001}_{-0.001}$ |
| $^{220}At$ | $^{216}Bi$ | $3^{[-]}$ | $(6^-, 7^-)$ | 4 | $6.077 \pm 0.018$ | $(2.976 \pm 0.774) \times 10^3$ | $0.0935 \pm 0.0421$ | 0.0161 | $14.441^{+4.138}_{-2.427} \times 10^3$ |
| $^{193}Rn$ | $^{189}Po$ | $[3/2^-]$ | $(5/2^-)$ | 2 | $8.040 \pm 0.012$ | $(1.150 \pm 0.270) \times 10^{-3}$ | $0.0330 \pm 0.0108$ | 0.0091 | $3.828^{+0.419}_{-0.354} \times 10^{-3}$ |
| $^{205}Rn$ | $^{201}Po$ | $5/2^-$ | $3/2^-$ | 2 | $6.390 \pm 0.050$ | $(6.918 \pm 0.424) \times 10^2$ | $0.0960 \pm 0.0488$ | 0.0079 | $77.117^{+41.839}_{-33.317} \times 10^2$ |
| $^{211}Rn$ | $^{207}Po$ | $1/2^-$ | $5/2^-$ | 2 | $5.965 \pm 0.001$ | $(1.927 \pm 0.146) \times 10^5$ | $0.0178 \pm 0.0016$ | 0.0029 | $11.749^{+0.253}_{-0.114} \times 10^5$ |
| $^{213}Rn$ | $^{209}Po$ | $[9/2^+]$ | $1/2^-$ | 5 | $8.243 \pm 0.005$ | $(1.950 \pm 0.010) \times 10^{-2}$ | $0.0055 \pm 0.0003$ | 0.0150 | $0.693^{+0.049}_{-0.015} \times 10^{-2}$ |
| $^{219}Rn$ | $^{215}Po$ | $5/2^+$ | $9/2^+$ | 2 | 6.946 | $3.960 \pm 0.010$ | $0.0198 \pm 0.0001$ | 0.0097 | 8.112 |
| $^{221}Rn$ | $^{217}Po$ | $7/2^+$ | $(9/2^+)$ | 2 | $6.162 \pm 0.002$ | $(7.030 \pm 0.456) \times 10^3$ | $0.0118 \pm 0.0010$ | 0.0104 | $8.000^{+0.125}_{-0.191} \times 10^3$ |
| $^{210}Fr$ | $^{206}At$ | $6^+$ | $(5)^+$ | 2 | $6.672 \pm 0.005$ | $(1.908 \pm 0.036) \times 10^2$ | $0.0342 \pm 0.0024$ | 0.0029 | $22.211^{+1.263}_{-1.009} \times 10^2$ |
| $^{212}Fr$ | $^{208}At$ | $5^+$ | $6^+$ | 2 | $6.529 \pm 0.002$ | $(1.200 \pm 0.036) \times 10^3$ | $0.0242 \pm 0.0013$ | 0.0013 | $22.762^{+0.680}_{-0.418} \times 10^3$ |
| $^{214}Fr$ | $^{210}At$ | $[1^-]$ | $(5)^+$ | 5 | $8.589 \pm 0.004$ | $(5.000 \pm 0.200) \times 10^{-3}$ | $0.0051 \pm 0.0002$ | 0.0115 | $2.218^{+0.018}_{-0.019} \times 10^{-3}$ |
| $^{220}Fr$ | $^{216}At$ | $1^+$ | $1^{(-)}$ | 1 | $6.801 \pm 0.002$ | $27.400 \pm 0.300$ | $0.0296 \pm 0.0008$ | 0.0053 | $156.792^{+0.304}_{-4.966}$ |
| $^{221}Fr$ | $^{217}At$ | $5/2^-$ | $9/2^-$ | 2 | $6.458 \pm 0.001$ | $286.620 \pm 0.780$ | $0.0425 \pm 0.0007$ | 0.0109 | $1121.900^{+13.423}_{-17.376}$ |
| $^{207}Ra$ | $^{203}Rn$ | $[5/2^-]$ | $[3/2^-]$ | 2 | $7.270 \pm 0.050$ | $1.605 \pm 0.209$ | $0.0654 \pm 0.0326$ | 0.0085 | $10.538^{+5.513}_{-3.551}$ |
| $^{213}Ra$ | $^{209}Rn$ | $1/2^-$ | $5/2^-$ | 2 | $6.862 \pm 0.002$ | $(1.638 \pm 0.030) \times 10^2$ | $0.0144 \pm 0.0005$ | 0.0032 | $7.215^{+0.191}_{-0.025} \times 10^2$ |
| $^{215}Ra$ | $^{211}Rn$ | $[9/2^+]$ | $1/2^-$ | 5 | $8.864 \pm 0.003$ | $(1.670 \pm 0.010) \times 10^{-3}$ | $0.0064 \pm 0.0001$ | 0.0158 | $0.677^{+0.011}_{-0.010} \times 10^{-3}$ |
| $^{219}Ra$ | $^{215}Rn$ | $(7/2)^+$ | $9/2^+$ | 2 | $8.138 \pm 0.003$ | $0.010 \pm 0.003$ | $0.0170 \pm 0.0054$ | 0.0092 | $0.017^{+0.000}_{-0.001}$ |
| $^{221}Ra$ | $^{217}Rn$ | $5/2^+$ | $9/2^+$ | 2 | $6.880 \pm 0.002$ | $28.000 \pm 2.000$ | $0.0259 \pm 0.0024$ | 0.0104 | $70.838^{+0.232}_{-2.684}$ |
| $^{223}Ra$ | $^{219}Rn$ | $3/2^+$ | $5/2^+$ | 2 | 5.979 | $(98.755 \pm 0.432) \times 10^4$ | 0.0017 | 0.0111 | $15.389 \times 10^4$ |
| $^{210}Ac$ | $^{206}Fr$ | $[7^+]$ | $(2^+, 3^+)$ | 4 | $7.610 \pm 0.050$ | $0.350 \pm 0.040$ | $0.1806 \pm 0.0862$ | 0.0098 | $5.594^{+2.810}_{-1.852}$ |
| $^{214}Ac$ | $^{210}Fr$ | $[5^+]$ | $6^+$ | 2 | $7.352 \pm 0.003$ | $8.200 \pm 0.200$ | $0.0129 \pm 0.0006$ | 0.0015 | $69.508^{+1.311}_{-1.967}$ |
| $^{216}Ac$ | $^{212}Fr$ | $(1^-)$ | $5^+$ | 5 | $9.235 \pm 0.006$ | $(4.400 \pm 0.160) \times 10^{-4}$ | $0.0059 \pm 0.0004$ | 0.0116 | $2.273^{+0.039}_{-0.121} \times 10^{-4}$ |
| $^{220}Ac$ | $^{216}Fr$ | $(3^-)$ | $(1^-)$ | 2 | $8.348 \pm 0.004$ | $(2.636 \pm 0.019) \times 10^{-2}$ | 0.0014 | 0.0079 | $0.480^{+0.015}_{-0.012} \times 10^{-2}$ |
| $^{223}Ac$ | $^{219}Fr$ | $(5/2^-)$ | $9/2^-$ | 2 | $6.783 \pm 0.001$ | $127.273 \pm 3.030$ | $0.0176 \pm 0.0007$ | 0.0113 | $197.258^{+3.077}_{-2.431}$ |
| $^{224}Ac$ | $^{220}Fr$ | $0^-$ | $1^+$ | 1 | $6.327 \pm 0.001$ | $(1.113 \pm 0.266) \times 10^5$ | $0.0008 \pm 0.0002$ | 0.0061 | $0.143^{+0.001}_{-0.002} \times 10^5$ |
| $^{225}Ac$ | $^{221}Fr$ | $[3/2^-]$ | $5/2^-$ | 2 | $5.935 \pm 0.001$ | $(857.088 \pm 0.259) \times 10^3$ | $0.0075 \pm 0.0001$ | 0.0119 | $540.738^{+9.430}_{-8.475} \times 10^3$ |
| $^{226}Ac$ | $^{222}Fr$ | $(1)^{[-]}$ | $2^-$ | 2 | $5.536 \pm 0.021$ | $(1.985 \pm 0.669) \times 10^9$ | $0.0003 \pm 0.0002$ | 0.0102 | $0.048^{+0.015}_{-0.011} \times 10^9$ |
| $^{209}Th$ | $^{205}Ra$ | $[5/2^-]$ | $(3/2^-)$ | 2 | $8.270 \pm 0.050$ | $(3.100 \pm 1.200) \times 10^{-3}$ | $0.0855 \pm 0.0546$ | 0.0089 | $21.244^{+8.709}_{-6.324} \times 10^{-3}$ |
| $^{211}Th$ | $^{207}Ra$ | $[5/2^-]$ | $[3/2^-]$ | 2 | $7.940 \pm 0.050$ | $0.048 \pm 0.020$ | $0.0761 \pm 0.0480$ | 0.0071 | $0.338^{+0.153}_{-0.068}$ |
| $^{215}Th$ | $^{211}Ra$ | $(1/2^-)$ | $5/2^-$ | 2 | $7.665 \pm 0.004$ | $1.200 \pm 0.200$ | $0.0207 \pm 0.0043$ | 0.0035 | $6.803^{+0.377}_{-0.184}$ |
| $^{217}Th$ | $^{213}Ra$ | $[9/2^+]$ | $1/2^-$ | 5 | $9.435 \pm 0.004$ | $(2.470 \pm 0.040) \times 10^{-4}$ | $0.0075 \pm 0.0003$ | 0.0161 | $1.153^{+0.020}_{-0.024} \times 10^{-4}$ |
| $^{221}Th$ | $^{217}Ra$ | $(7/2^+)$ | $(9/2^+)$ | 2 | $8.626 \pm 0.004$ | $(1.680 \pm 0.060) \times 10^{-3}$ | $0.0075 \pm 0.0005$ | 0.0096 | $1.311^{+0.054}_{-0.031} \times 10^{-3}$ |
| $^{223}Th$ | $^{219}Ra$ | $(5/2)^+$ | $(7/2)^+$ | 2 | $7.567 \pm 0.004$ | $0.600 \pm 0.020$ | $0.0129 \pm 0.0008$ | 0.0108 | $0.716^{+0.022}_{-0.023}$ |
| $^{225}Th$ | $^{221}Ra$ | $(3/2^+)$ | $5/2^+$ | 2 | $6.921 \pm 0.002$ | $525.000 \pm 2.400$ | 0.0019 | 0.0116 | $85.584^{+1.402}_{-1.627}$ |
| $^{227}Th$ | $^{223}Ra$ | $1/2^+$ | $3/2^+$ | 2 | 6.147 | $(161.395 \pm 0.778) \times 10^4$ | 0.0006 | 0.0118 | $7.803 \times 10^4$ |
| $^{229}Th$ | $^{225}Ra$ | $5/2^+$ | $1/2^+$ | 2 | $5.168 \pm 0.001$ | $(2.503 \pm 0.017) \times 10^{11}$ | 0.0007 | 0.0114 | $0.161^{+0.002}_{-0.004} \times 10^{11}$ |
| $^{224}Pa$ | $^{220}Ac$ | $[5^-]$ | $(3^-)$ | 2 | $7.694 \pm 0.004$ | $0.844 \pm 0.019$ | $0.0072 \pm 0.0004$ | 0.0093 | $0.655^{+0.021}_{-0.020}$ |
| $^{225}Pa$ | $^{221}Ac$ | $[5/2^-]$ | $[9/2^-]$ | 2 | $7.390 \pm 0.050$ | $1.700 \pm 0.200$ | $0.0330 \pm 0.0161$ | 0.0115 | $4.241^{+2.166}_{-1.454}$ |
| $^{228}Pa$ | $^{224}Ac$ | $3^+$ | $0^-$ | 3 | $6.265 \pm 0.002$ | $(4.018 \pm 0.582) \times 10^6$ | 0.0003 | 0.0144 | $0.071^{+0.002}_{-0.001} \times 10^6$ |
| $^{229}Pa$ | $^{225}Ac$ | $(5/2^+)$ | $[3/2^-]$ | 1 | $5.835 \pm 0.004$ | $(2.739 \pm 0.375) \times 10^7$ | $0.0022 \pm 0.0004$ | 0.0075 | $0.765^{+0.041}_{-0.034} \times 10^7$ |
| $^{230}Pa$ | $^{226}Ac$ | $(2^-)$ | $(1)^{[-]}$ | 2 | $5.439 \pm 0.001$ | $(4.707 \pm 0.282) \times 10^{10}$ | 0.0004 | 0.0100 | $0.167^{+0.002}_{-0.002} \times 10^{10}$ |
| $^{217}U$ | $^{213}Th$ | $[1/2^-]$ | $[5/2^-]$ | 2 | $8.430 \pm 0.070$ | $0.016^{+0.021}_{-0.006}$ [30] | $0.0413 \pm 0.0336$ | 0.0036 | $0.129^{+0.081}_{-0.049}$ |
| $^{219}U$ | $^{215}Th$ | $[9/2^+]$ | $(1/2^-)$ | 5 | $9.940 \pm 0.050$ | $(5.500 \pm 2.500) \times 10^{-5}$ | $0.0145 \pm 0.0094$ | 0.0160 | $3.379^{+1.091}_{-0.791} \times 10^{-5}$ |
| $^{223}U$ | $^{219}Th$ | $[7/2^+]$ | $[9/2^+]$ | 2 | $8.940 \pm 0.050$ | $(2.100 \pm 0.800) \times 10^{-5}$ | $0.5986 \pm 0.3679$ | 0.0097 | $94.670^{+35.152}_{-25.529} \times 10^{-5}$ |
| $^{225}U$ | $^{221}Th$ | $[5/2^+]$ | $(7/2^+)$ | 2 | $8.015 \pm 0.007$ | $0.061 \pm 0.004$ | $0.0240 \pm 0.0029$ | 0.0109 | $0.133^{+0.008}_{-0.007}$ |
| $^{227}U$ | $^{223}Th$ | $(3/2^+)$ | $(5/2)^+$ | 2 | $7.211 \pm 0.014$ | $66.000 \pm 6.000$ | $0.0048 \pm 0.0010$ | 0.0117 | $26.775^{+3.244}_{-3.244}$ |
| $^{231}U$ | $^{227}Th$ | $(5/2)^{[+]}$ | $1/2^+$ | 2 | $5.576 \pm 0.002$ | $(9.734 \pm 2.650) \times 10^9$ | $0.0026 \pm 0.0009$ | 0.0116 | $1.834^{+0.330}_{-0.068} \times 10^9$ |
| $^{235}U$ | $^{231}Th$ | $7/2^-$ | $5/2^+$ | 1 | $4.678 \pm 0.001$ | $(2.221 \pm 0.002) \times 10^{16}$ | 0.0003 | 0.0057 | $0.116^{+0.002}_{-0.003} \times 10^{16}$ |
| $^{227}Np$ | $^{223}Pa$ | $[5/2^-]$ | $[9/2^-]$ | 2 | $7.816 \pm 0.014$ | $0.510 \pm 0.060$ | $0.0147 \pm 0.0033$ | 0.0114 | $0.634^{+0.076}_{-0.064}$ |
| $^{229}Np$ | $^{225}Pa$ | $[5/2^+]$ | $[5/2^-]$ | 1 | $7.010 \pm 0.050$ | $(2.400 \pm 0.108) \times 10^{2}$ | $0.0103 \pm 0.0048$ | 0.0074 | $2.933^{+1.744}_{-1.072} \times 10^{2}$ |
| $^{231}Np$ | $^{227}Pa$ | $(5/2)^{[+]}$ | $(5/2^-)$ | 1 | $6.370 \pm 0.050$ | $(1.956 \pm 0.984) \times 10^{5}$ | $0.0388 \pm 0.0309$ | 0.0075 | $5.309^{+3.793}_{-2.197} \times 10^{5}$ |
| $^{235}Np$ | $^{231}Pa$ | $5/2^+$ | $3/2^-$ | 1 | $5.194 \pm 0.002$ | $(1.320 \pm 0.070) \times 10^{12}$ | $0.0055 \pm 0.0005$ | 0.0064 | $1.126^{+0.046}_{-0.025} \times 10^{12}$ |
| $^{236}Np$ | $^{232}Pa$ | $(6^-)$ | $(2^-)$ | 4 | $5.010 \pm 0.050$ | $(3.245 \pm 0.910) \times 10^{15}$ | $0.0001 \pm 0.0001$ | 0.0135 | $0.019^{+0.010}_{-0.011} \times 10^{15}$ |
| $^{237}Np$ | $^{233}Pa$ | $5/2^+$ | $3/2^-$ | 1 | $4.959 \pm 0.001$ | $(6.766 \pm 0.022) \times 10^{13}$ | 0.0031 | 0.0055 | $3.841^{+0.078}_{-0.005} \times 10^{13}$ |
| $^{229}Pu$ | $^{225}U$ | $[3/2^+]$ | $[5/2^+]$ | 2 | $7.590 \pm 0.050$ | $91.000 \pm 26.000$ | $0.0083 \pm 0.0051$ | 0.0116 | $50.039^{+25.653}_{-17.556}$ |
| $^{233}Pu$ | $^{229}U$ | $(5/2^+)$ | $(3/2^+)$ | 2 | $6.420 \pm 0.050$ | $(1.275 \pm 0.055) \times 10^{6}$ | $0.0067 \pm 0.0035$ | 0.0114 | $0.634^{+0.449}_{-0.259} \times 10^{6}$ |
| $^{237}Pu$ | $^{233}U$ | $7/2^-$ | $5/2^+$ | 1 | $5.748 \pm 0.002$ | $(9.476 \pm 0.911) \times 10^{10}$ | 0.0001 | 0.0057 | $0.180^{+0.003}_{-0.005} \times 10^{10}$ |
| $^{239}Pu$ | $^{235}U$ | $1/2^+$ | $7/2^-$ | 3 | 5.245 | $(7.608 \pm 0.009) \times 10^{11}$ | 0.0318 | 0.0101 | $23.904 \times 10^{11}$ |
| $^{241}Pu$ | $^{237}U$ | $5/2^+$ | $1/2^+$ | 2 | $5.140 \pm 0.001$ | $(1.840 \pm 0.016) \times 10^{13}$ | $0.0038 \pm 0.0001$ | 0.0057 | $1.232^{+0.008}_{-0.033} \times 10^{13}$ |
| $^{235}Am$ | $^{231}Np$ | $[5/2^-]$ | $(5/2)^{[+]}$ | 1 | $6.576 \pm 0.013$ | $(1.400 \pm 0.236) \times 10^{5}$ [32] | $0.0189 \pm 0.0057$ | 0.0069 | $3.596^{+0.545}_{-0.458} \times 10^{5}$ |
| $^{239}Am$ | $^{235}Np$ | $(5/2)^-$ | $5/2^+$ | 1 | $5.922 \pm 0.001$ | $(4.331 \pm 0.469) \times 10^{8}$ | $0.0087 \pm 0.0011$ | 0.0053 | $6.952^{+0.148}_{-0.101} \times 10^{8}$ |
| $^{240}Am$ | $^{236}Np$ | $(3^-)$ | $(6^-)$ | 4 | $5.710 \pm 0.050$ | $(1.116 \pm 0.417) \times 10^{11}$ | $0.0012 \pm 0.0009$ | 0.0101 | $0.076^{+0.074}_{-0.035} \times 10^{11}$ |
| $^{241}Am$ | $^{237}Np$ | $5/2^-$ | $5/2^+$ | 1 | 5.638 | $(1.365 \pm 0.002) \times 10^{10}$ | 0.0105 | 0.0043 | $3.330 \times 10^{10}$ |
| $^{243}Am$ | $^{239}Np$ | $5/2^-$ | $5/2^+$ | 1 | $5.439 \pm 0.001$ | $(2.326 \pm 0.013) \times 10^{11}$ | $0.0086 \pm 0.0002$ | 0.0033 | $6.045^{+0.116}_{-0.037} \times 10^{11}$ |
| $^{239}Cm$ | $^{235}Pu$ | $(7/2^-)$ | $(5/2^+)$ | 1 | $6.540 \pm 0.050$ | $(1.585 \pm 0.590) \times 10^{8}$ | $0.0001 \pm 0.0001$ | 0.0055 | $0.018^{+0.013}_{-0.008} \times 10^{8}$ |
| $^{241}Cm$ | $^{237}Pu$ | $1/2^+$ | $7/2^-$ | 3 | $6.185 \pm 0.001$ | $(2.864 \pm 0.304) \times 10^{8}$ | $0.0048 \pm 0.0006$ | 0.0097 | $1.400^{+0.028}_{-0.024} \times 10^{8}$ |
| $^{243}Cm$ | $^{239}Pu$ | $5/2^+$ | $1/2^+$ | 2 | $6.169 \pm 0.001$ | $(9.183 \pm 0.032) \times 10^{8}$ | 0.0009 | 0.0055 | $1.546^{+0.004}_{-0.007} \times 10^{8}$ |
| $^{245}Cm$ | $^{241}Pu$ | $7/2^+$ | $5/2^+$ | 2 | $5.623 \pm 0.001$ | $(2.658 \pm 0.023) \times 10^{11}$ | $0.0033 \pm 0.0001$ | 0.0039 | $2.290^{+0.009}_{-0.073} \times 10^{11}$ |
| $^{247}Cm$ | $^{243}Pu$ | $9/2^-$ | $7/2^+$ | 1 | $5.354 \pm 0.003$ | $(4.923 \pm 0.158) \times 10^{14}$ | 0.0001 | 0.0017 | $0.152^{+0.004}_{-0.008} \times 10^{14}$ |
| $^{243}Bk$ | $^{239}Am$ | $[3/2^-]$ | $(5/2)^-$ | 2 | $6.874 \pm 0.004$ | $(1.080 \pm 0.048) \times 10^{7}$ | 0.0001 | 0.0065 | $0.022^{+0.001}_{-0.001} \times 10^{7}$ |
| $^{244}Bk$ | $^{240}Am$ | $[4^-]$ | $(3^-)$ | 2 | $6.779 \pm 0.004$ | $(3.540 \pm 1.860) \times 10^{8}$ | $0.00001 \pm 0.00001$ | 0.0038 | $0.008 \times 10^{8}$ |
| $^{245}Bk$ | $^{241}Am$ | $3/2^-$ | $5/2^-$ | 2 | $6.455 \pm 0.001$ | $(3.591 \pm 0.321) \times 10^{8}$ | 0.0002 | 0.0049 | $0.155^{+0.000}_{-0.005} \times 10^{8}$ |
| $^{247}Bk$ | $^{243}Am$ | $(3/2^-)$ | $5/2^-$ | 2 | $5.890 \pm 0.005$ | $(4.355 \pm 0.789) \times 10^{10}$ | $0.0019 \pm 0.0005$ | 0.0017 | $4.854^{+0.294}_{-0.331} \times 10^{10}$ |
| $^{249}Bk$ | $^{245}Am$ | $7/2^+$ | $(5/2)^+$ | 2 | $5.523 \pm 0.002$ | $(1.974 \pm 0.133) \times 10^{12}$ | $0.0055 \pm 0.0005$ | 0.0026 | $4.148^{+0.092}_{-0.127} \times 10^{12}$ |
| $^{237}Cf$ | $^{233}Cm$ | $[5/2^+]$ | $[3/2^+]$ | 2 | $8.220 \pm 0.050$ | $0.800 \pm 0.200$ | $0.0561 \pm 0.0315$ | 0.0103 | $3.514^{+1.597}_{-1.127}$ |
| $^{247}Cf$ | $^{243}Cm$ | $[7/2^+]$ | $5/2^+$ | 2 | $6.495 \pm 0.015$ | $(3.270 \pm 0.498) \times 10^{7}$ | $0.0049 \pm 0.0015$ | 0.0036 | $4.225^{+0.765}_{-0.622} \times 10^{7}$ |
| $^{249}Cf$ | $^{245}Cm$ | $9/2^-$ | $7/2^+$ | 1 | $6.296 \pm 0.001$ | $(1.108 \pm 0.006) \times 10^{10}$ | 0.0001 | 0.0016 | $0.075^{+0.001}_{-0.000} \times 10^{10}$ |
| $^{251}Cf$ | $^{247}Cm$ | $1/2^+$ | $9/2^-$ | 5 | $6.177 \pm 0.001$ | $(2.840 \pm 0.126) \times 10^{10}$ | $0.0027 \pm 0.0002$ | 0.0067 | $1.156^{+0.018}_{-0.012} \times 10^{10}$ |
| $^{253}Cf$ | $^{249}Cm$ | $(7/2^+)$ | $[1/2^+]$ | 4 | $6.126 \pm 0.004$ | $(5.051 \pm 0.674) \times 10^{8}$ | $0.1017 \pm 0.0158$ | 0.0070 | $72.715^{+1.309}_{-1.957} \times 10^{8}$ |
| $^{245}Es$ | $^{241}Bk$ | $(3/2^-)$ | $(7/2^+)$ | 3 | $7.909 \pm 0.003$ | $(1.800 \pm 0.600) \times 10^{2}$ | $0.0096 \pm 0.0034$ | 0.0080 | $1.907^{+0.045}_{-0.037} \times 10^{2}$ |
| $^{246}Es$ | $^{242}Bk$ | $[4^-]$ | $[2^-]$ | 2 | $7.740 \pm 0.100$ | $(4.758 \pm 1.168) \times 10^{3}$ | $0.0013 \pm 0.0010$ | 0.0034 | $1.048^{+1.414}_{-0.596} \times 10^{3}$ |
| $^{252}Es$ | $^{248}Bk$ | $(4^+)$ | $[6^+]$ | 2 | $6.790 \pm 0.050$ | $(4.075 \pm 0.016) \times 10^{7}$ | $0.0007 \pm 0.0003$ | 0.0017 | $1.550^{+0.983}_{-0.667} \times 10^{7}$ |
| $^{254}Es$ | $^{250}Bk$ | $(7^+)$ | $2^-$ | 5 | $6.616 \pm 0.002$ | $(2.382 \pm 0.004) \times 10^{7}$ | $0.0596 \pm 0.0013$ | 0.0052 | $27.470^{+0.424}_{-0.637} \times 10^{7}$ |
| $^{255}Es$ | $^{251}Bk$ | $(7/2^+)$ | $(3/2^-)$ | 3 | $6.436 \pm 0.001$ | $(4.316 \pm 0.345) \times 10^{7}$ | $0.0420 \pm 0.0036$ | 0.0056 | $32.031^{+0.018}_{-0.319} \times 10^{7}$ |
| $^{243}Fm$ | $^{239}Cf$ | $[7/2^-]$ | $[5/2^+]$ | 1 | $8.690 \pm 0.050$ | $0.231 \pm 0.009$ | $0.0205 \pm 0.0077$ | 0.0047 | $0.935^{+0.400}_{-0.282}$ |
| $^{245}Fm$ | $^{241}Cf$ | $[1/2^+]$ | $[7/2^-]$ | 3 | $8.440 \pm 0.100$ | $4.200 \pm 1.300$ | $0.0253 \pm 0.0199$ | 0.0082 | $7.525^{+8.438}_{-3.937}$ |
| $^{247}Fm$ | $^{243}Cf$ | $[7/2^+]$ | $[1/2^+]$ | 4 | $8.258 \pm 0.010$ | $31.000 \pm 1.000$ | $0.0135 \pm 0.0015$ | 0.0079 | $53.246^{+4.331}_{-4.204}$ |
| $^{249}Fm$ | $^{245}Cf$ | $(7/2^+)$ | $1/2^+$ | 4 | $7.709 \pm 0.006$ | $(5.482 \pm 2.768) \times 10^{2}$ [33] | $0.0936 \pm 0.0506$ | 0.0054 | $68.780^{+3.110}_{-3.579} \times 10^{2}$ |
| $^{251}Fm$ | $^{247}Cf$ | $(9/2^-)$ | $[7/2^+]$ | 1 | $7.425 \pm 0.002$ | $(1.066 \pm 0.087) \times 10^{6}$ | 0.0001 | 0.0015 | $0.062^{+0.001}_{-0.002} \times 10^{6}$ |
| $^{253}Fm$ | $^{249}Cf$ | $(1/2)^+$ | $9/2^-$ | 5 | $7.198 \pm 0.003$ | $(2.182 \pm 0.268) \times 10^{6}$ | $0.0050 \pm 0.0007$ | 0.0061 | $1.767^{+0.040}_{-0.046} \times 10^{6}$ |
| $^{255}Fm$ | $^{251}Cf$ | $7/2^+$ | $1/2^+$ | 4 | $7.240 \pm 0.002$ | $(7.225 \pm 0.025) \times 10^{4}$ | $0.0362 \pm 0.0008$ | 0.0063 | $42.000^{+0.238}_{-1.253} \times 10^{4}$ |
| $^{257}Fm$ | $^{253}Cf$ | $(9/2^+)$ | $(7/2^+)$ | 2 | $6.864 \pm 0.001$ | $(8.683 \pm 0.017) \times 10^6$ | $0.0040 \pm 0.0001$ | 0.0045 | $7.596^{+0.209}_{-0.002} \times 10^6$ |
| $^{247}Md$ | $^{243}Es$ | $(7/2^-)$ | $(7/2^+)$ | 1 | $8.764 \pm 0.010$ | $1.190 \pm 0.090$ | $0.0051 \pm 0.0008$ | 0.0033 | $1.808^{+0.133}_{-0.132}$ |
| $^{249}Md$ | $^{245}Es$ | $(7/2^-)$ | $(3/2^-)$ | 2 | $8.441 \pm 0.018$ | $23.400 \pm 2.400$ | $0.0034 \pm 0.0008$ | 0.0039 | $19.643^{+2.841}_{-2.402}$ |
| $^{251}Md$ | $^{247}Es$ | $(7/2^-)$ | $(7/2^+)$ | 1 | $7.963 \pm 0.004$ | $(2.565 \pm 0.395) \times 10^3$ | $0.0009 \pm 0.0002$ | 0.0008 | $3.021^{+0.078}_{-0.095} \times 10^3$ |
| $^{255}Md$ | $^{251}Es$ | $(7/2^-)$ | $(3/2^-)$ | 2 | $7.906 \pm 0.003$ | $(2.200 \pm 0.700) \times 10^4$ | $0.0002 \pm 0.0001$ | 0.0029 | $0.159^{+0.005}_{-0.005} \times 10^4$ |
| $^{256}Md$ | $^{252}Es$ | $(1^-)$ | $(4^+)$ | 3 | $7.856 \pm 0.016$ [34] | $(5.061 \pm 0.516) \times 10^4$ [34] | $0.0003 \pm 0.0001$ | 0.0028 | $0.449^{+0.071}_{-0.054} \times 10^4$ |
| $^{257}Md$ | $^{253}Es$ | $(7/2^-)$ | $7/2^+$ | 1 | $7.558 \pm 0.001$ | $(1.383 \pm 0.289) \times 10^5$ | $0.0005 \pm 0.0001$ | 0.0024 | $0.302^{+0.002}_{-0.000} \times 10^5$ |
| $^{258}Md$ | $^{254}Es$ | $(8^-)$ | $(7^+)$ | 1 | $7.271 \pm 0.002$ | $(4.450 \pm 0.026) \times 10^6$ | 0.0003 | 0.0019 | $0.610^{+0.010}_{-0.006} \times 10^6$ |
| $^{253}No$ | $^{249}Fm$ | $(9/2^-)$ | $(7/2^+)$ | 1 | $8.414 \pm 0.004$ | $93.600 \pm 1.200$ | $0.0018 \pm 0.0001$ | 0.0010 | $169.264^{+6.118}_{-5.098}$ |
| $^{255}No$ | $^{251}Fm$ | $(1/2^+)$ | $(9/2^-)$ | 5 | $8.428 \pm 0.003$ | $(2.112 \pm 0.108) \times 10^2$ | $0.0070 \pm 0.0005$ | 0.0040 | $3.707^{+0.089}_{-0.081} \times 10^2$ |
| $^{257}No$ | $^{253}Fm$ | $(7/2^+)$ | $(1/2)^+$ | 4 | $8.477 \pm 0.006$ | $24.500 \pm 0.500$ | $0.0183 \pm 0.0012$ | 0.0042 | $106.619^{+5.776}_{-3.610}$ |
| $^{259}No$ | $^{255}Fm$ | $[9/2^+]$ | $7/2^+$ | 2 | $7.858 \pm 0.005$ [35] | $(3.480 \pm 0.300) \times 10^3$ [35] | $0.0069 \pm 0.0009$ | 0.0030 | $7.834^{+0.332}_{-0.279} \times 10^3$ |
| $^{255}Lr$ | $^{251}Md$ | $(1/2^-)$ | $(7/2^-)$ | 4 | $8.556 \pm 0.007$ | $31.100 \pm 1.100$ | $0.0218 \pm 0.0018$ | 0.0035 | $194.067^{+7.706}_{-9.989}$ |
| $^{257}Lr$ | $^{253}Md$ | $(1/2^-)$ | $(7/2^-)$ | 4 | $9.010 \pm 0.030$ | $6.000 \pm 0.400$ | $0.0036 \pm 0.0009$ | 0.0046 | $4.404^{+0.978}_{-0.758}$ |
| $^{259}Lr$ | $^{255}Md$ | $[1/2^-]$ | $(7/2^-)$ | 4 | $8.580 \pm 0.070$ | $7.964 \pm 0.589$ | $0.0934 \pm 0.0509$ | 0.0058 | $107.651^{+76.795}_{-44.551}$ |
| $^{255}Rf$ | $^{251}No$ | $(9/2^-)$ | $(7/2^+)$ | 1 | $9.055 \pm 0.004$ | $1.660 \pm 0.070$ | $0.0057 \pm 0.0004$ | 0.0012 | $8.042^{+0.245}_{-0.216}$ |
| $^{257}Rf$ | $^{253}No$ | $(1/2^+)$ | $(9/2^-)$ | 5 | $9.083 \pm 0.008$ | $4.820 \pm 0.130$ | $0.0166 \pm 0.0014$ | 0.0047 | $16.910^{+1.041}_{-0.962}$ |
| $^{259}Rf$ | $^{255}No$ | $[7/2^+]$ | $(1/2^+)$ | 4 | $9.130 \pm 0.070$ | $2.630 \pm 0.260$ | $0.0104 \pm 0.0054$ | 0.0050 | $4.655^{+2.894}_{-1.759}$ |
| $^{261}Rf$ | $^{257}No$ | $[3/2^+]$ | $(7/2^+)$ | 2 | $8.650 \pm 0.050$ | $8.831 \pm 3.074$ | $0.0335 \pm 0.0211$ | 0.0036 | $59.512^{+27.403}_{-18.546}$ |
| $^{257}Db$ | $^{253}Lr$ | $(9/2^+)$ | $(7/2^-)$ | 1 | $9.206 \pm 0.020$ | $2.300 \pm 0.200$ | $0.0034 \pm 0.0008$ | 0.0014 | $5.361^{+0.734}_{-0.755}$ |
| $^{259}Db$ | $^{255}Lr$ | $[9/2^+]$ | $(1/2^-)$ | 5 | $9.620 \pm 0.050$ | $0.510 \pm 0.160$ [35] | $0.0108 \pm 0.0062$ | 0.0061 | $0.704^{+0.276}_{-0.193}$ |
| $^{259}Sg$ | $^{255}Rf$ | $[1/2^+]$ | $(9/2^-)$ | 5 | $9.804 \pm 0.021$ | $0.280 \pm 0.050$ | $0.0124 \pm 0.0038$ | 0.0054 | $0.597^{+0.088}_{-0.075}$ |
| $^{261}Sg$ | $^{257}Rf$ | $(3/2^+)$ | $(1/2^+)$ | 2 | $9.714 \pm 0.015$ | $0.183 \pm 0.005$ | $0.0042 \pm 0.0005$ | 0.0034 | $0.220^{+0.023}_{-0.019}$ |
| $^{265}Sg$ | $^{261}Rf$ | $[9/2^+]$ | $[3/2^+]$ | 4 | $8.823 \pm 0.051$ [36] | $16.500 \pm 5.500$ [36] | $0.1112 \pm 0.0701$ | 0.0078 | $169.980^{+85.438}_{-54.055}$ |
| $^{261}Bh$ | $^{257}Db$ | $(5/2^-)$ | $(9/2^+)$ | 3 | $10.500 \pm 0.050$ | $(1.28 \pm 0.320) \times 10^{-2}$ | $0.0021 \pm 0.0010$ | 0.0044 | $0.500^{+0.167}_{-0.123} \times 10^{-2}$ |
| $^{263}Hs$ | $^{259}Sg$ | $[7/2^+]$ | $[1/2^+]$ | 4 | $10.730 \pm 0.050$ | $(7.600 \pm 0.400) \times 10^{-4}$ | $0.0337 \pm 0.0110$ | 0.0063 | $38.544^{+12.347}_{-98.178} \times 10^{-4}$ |
| $^{267}Hs$ | $^{263}Sg$ | $[5/2^+]$ | $[7/2^+]$ | 2 | $10.037 \pm 0.013$ | $(5.500 \pm 1.100) \times 10^{-2}$ | $0.0097 \pm 0.0027$ | 0.0053 | $9.581^{+0.754}_{-0.767} \times 10^{-2}$ |
| $^{267}Ds$ | $^{263}Hs$ | $[9/2^+]$ | $[7/2^+]$ | 2 | $11.780 \pm 0.050$ | $(1.000 \pm 0.800) \times 10^{-5}$ | $0.0277 \pm 0.0241$ | 0.0050 | $1.613^{+0.460}_{-0.337} \times 10^{-5}$ |
| $^{269}Ds$ | $^{265}Hs$ | $[9/2^+]$ | $[3/2^+]$ | 4 | $11.509 \pm 0.030$ | $(2.300 \pm 1.100) \times 10^{-4}$ | $0.0073 \pm 0.0043$ | 0.0093 | $1.279^{+0.217}_{-0.187} \times 10^{-4}$ |
| $^{271}Ds$ | $^{267}Hs$ | $[13/2^-]$ | $[5/2^+]$ | 5 | $10.870 \pm 0.018$ | $0.090 \pm 0.040$ | $0.0018 \pm 0.0009$ | 0.0120 | $0.010^{+0.001}_{-0.001}$ |
| $^{273}Ds$ | $^{269}Hs$ | $[13/2^-]$ | $[9/2^+]$ | 3 | $11.370 \pm 0.050$ | $(2.400 \pm 1.200) \times 10^{-4}$ | $0.0102 \pm 0.0069$ | 0.0084 | $1.872^{+0.577}_{-0.431} \times 10^{-4}$ |
| $^{277}Ds$ | $^{273}Hs$ | $[11/2^+]$ | $[3/2^+]$ | 4 | $10.840 \pm 0.110$ | $0.022 \pm 0.017$ | $0.0098 \pm 0.0091$ | 0.0092 | $0.006^{+0.005}_{-0.003}$ |
| $^{277}Cn$ | $^{273}Ds$ | $[3/2^+]$ | $[13/2^-]$ | 5 | $11.620 \pm 0.050$ | $(9.900 \pm 4.900) \times 10^{-4}$ | $0.0128 \pm 0.0085$ | 0.0125 | $6.520^{+2.007}_{-1.488} \times 10^{-4}$ |
| $^{281}Cn$ | $^{277}Ds$ | $[3/2^+]$ | $[11/2^+]$ | 4 | $10.460 \pm 0.050$ | $0.370 \pm 0.290$ | $0.0289 \pm 0.0253$ | 0.0084 | $0.378^{+0.137}_{-0.100}$ |
| $^{285}Cn$ | $^{281}Ds$ | $[5/2^+]$ | $[3/2^+]$ | 2 | $9.320 \pm 0.050$ | $32.000 \pm 9.000$ | $0.1068 \pm 0.0618$ | 0.0033 | $831.107^{+346.168}_{-271.468}$ |

**Table II**:

| $Z_0$ | $N_0$ | $Z_c$ | $N_c$ | $A$ |
|---|---|---|---|---|
| 50 | 82 | 8 | 8 | 0.110 |
| 70 | 82 | 6 | 8 | 0.067 |
| | 102 | | 12 | 0.062 |
| 82 | 82 | 12 | 8 | 0.073 |
| | 102 | | 12 | 0.078 |
| | 126 | | 12 | 0.105 |
| | 150 | | 14 | 0.087 |
| 102 | 126 | 14 | 12 | 0.108 |
| | 150 | | 14 | 0.103 |